\documentclass[english,aps,prb,reprint,twocolumn]{revtex4}
\usepackage[T1]{fontenc}
\usepackage[latin9]{inputenc}
\usepackage{color}
\usepackage{units}
\usepackage{amstext}
\usepackage{graphicx}
\usepackage{subscript}

\makeatletter
\@ifundefined{textcolor}{}
{%
 \definecolor{BLACK}{gray}{0}
 \definecolor{WHITE}{gray}{1}
 \definecolor{RED}{rgb}{1,0,0}
 \definecolor{GREEN}{rgb}{0,1,0}
 \definecolor{BLUE}{rgb}{0,0,1}
 \definecolor{CYAN}{cmyk}{1,0,0,0}
 \definecolor{MAGENTA}{cmyk}{0,1,0,0}
 \definecolor{YELLOW}{cmyk}{0,0,1,0}
}

\renewcommand{\fnum@figure}{\textbf{FIG. \thefigure}}

\definecolor{green}{RGB}{0, 180, 0}
\definecolor{cyan}{RGB}{0, 180, 180}
\definecolor{yellow}{RGB}{211,211,0}

\newcommand{\putat}[3]{\begin{picture}(0,0)(0,0)\put(#1,#2){#3}\end{picture}}

\usepackage{rotating}

\makeatother

\usepackage{babel}
\begin{document}

\title{Mechanically induced metal--insulator transition in carbyne}

\author{Vasilii I. Artyukhov}

\author{Mingjie Liu}

\affiliation{Department of Mechanical Engineering and Materials Science, Rice
University, Houston, Texas 77005}

\author{Boris I. Yakobson}

\affiliation{Department of Mechanical Engineering and Materials Science, Department
of Chemistry and Smalley Institute for Nanoscale Science and Technology,
Rice University, Houston, Texas 77005, USA }

\email{biy@rice.edu}

\begin{abstract}
First-principles calculations for carbyne under strain predict that
the Peierls transition from symmetric cumulene to broken-symmetry
polyyne structure is enhanced as the material is stretched. Interpretation
within a simple and instructive analytical model suggests that this
behavior is valid for arbitrary 1D metals. Further, numerical calculations
of the anharmonic quantum vibrational structure of carbyne show that
zero-point atomic vibrations alone eliminate the Peierls distortion
in a mechanically free chain, preserving the cumulene symmetry. The
emergence and increase of Peierls dimerization under tension then
implies a qualitative transition between the two forms, which our
computations place around 3\% strain. Thus, zero-point vibrations
and mechanical strain jointly produce a change in symmetry resulting
in the transition from metallic to insulating state. In any practical
realization, it is important that the effect is also chemically modulated
by the choice of terminating groups. Our findings are promising for
applications such as electromechanical switching and band gap tuning
via strain, and besides carbyne itself, they directly extend to numerous
other systems that show Peierls distortion. 
\end{abstract}
\maketitle
Carbyne---the linear allotrope of carbon---is perhaps one of the
most unusual materials due to its ultimate one-atom thinness. Although
carbyne is elusively hard to prepare and has been perceived as an
exotic or even completely fictitious material, the development of
methods to synthesize carbon chains proceeds at a steady rate, with
input from both experiments and theory.\cite{2005cataldopolyynes,2009jinderiving,2010hobiformation,2011erdoganengineering,2013casillasnewinsights,1997yakobsonhighstrain,2003troianidirectobservation}
Among the most notable recent achievements, chains with length of
up to 44 atoms \cite{2010chalifouxsynthesis} and such complex molecular
machines as carbyne-based rotaxanes \cite{2012movsisyansynthesis,2012weisbachanew}
have been synthesized. This progress is driven by carbyne's attractive
physical properties such as unusual electrical transport \cite{2008khoonegative,2010zanolliquantum}
and intriguing mechanics,\cite{2012liucarbynes} or its large specific
area.\cite{2011sorokincalciumdecorated} Accordingly, a better theoretical
understanding of this material is becoming more and more relevant.\cite{2012liucarbynes}

It has long ago been established by the quantum chemistry community
\cite{1978kerteszabinitio} that carbyne undergoes the Peierls transition
\cite{1930peierlszurtheorie,2001peierlsquantum,1954frohlichonthe,1987kennedyproofof}
that converts it from the cumulene (=C=C=)\textsubscript{\emph{n}}
to the polyyne (--C$\equiv$C--)\textsubscript{\emph{n}} form. Later
it has been suggested that the zero-point vibrations (ZPV) may substantially
affect the Peierls instability \cite{1992mckenzieeffectof} and even
completely eliminate the distortion in carbyne.\cite{2005tongayatomicand}
As the symmetric and broken-symmetry forms have very distinct electronic
properties (metallic and insulating, respectively), this issue becomes
crucial from both the fundamental physicochemical perspective and
for applications in 1D conducting systems.

\begin{figure}
~\hfill \includegraphics[width=0.9\columnwidth]{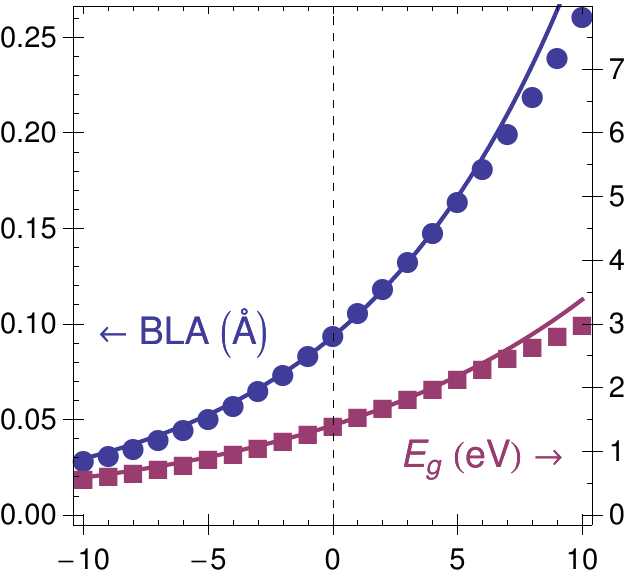}\putat{-185}{102}{\includegraphics[width=0.37\columnwidth]{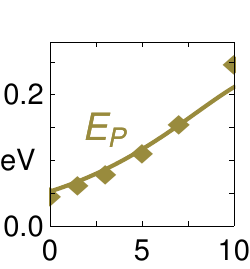}}\putat{-120}{-8}{\large{$\textsf{strain (\%)}$}}\hfill~\smallskip

\caption{\textbf{Bond length alternation (}\textbf{\textcolor{blue}{BLA}}\textbf{),
band gap (}\textbf{\textcolor{magenta}{\emph{E\textsubscript{\textbf{\emph{g}}}}}}\textbf{),
and Peierls barrier (}\textbf{\textcolor{yellow}{\emph{E\textsubscript{\textbf{\emph{P}}}}}}\textbf{,
inset) as a function of strain.} Points, DFT calculation results.
Lines, fitting based on the analytical model. The fitting is performed
in the {[}--5\%,~5\%{]} strain range assuming that the lattice stiffness
decreases linearly with strain, and is done independently for BLA
and \emph{E}\textsubscript{\emph{g}}. The \emph{E\textsubscript{\emph{P}}}
line in the inset is computed based on the BLA fit.\label{fig:BLA+gap}}
\end{figure}

A whole new dimension is added to the situation by the unusual effects
of stretching on carbyne that we have recently found through first-principles
calculations \cite{2012liucarbynes} (also observed experimentally
after the present study had been completed.\cite{2013cretuelectrical})
Specifically, stretching increases the bond length alternation (BLA,
defined as the difference between the long and short bonds) and the
band gap. Although the effect of strain on the band gap is well-known
in semiconductors, the present case is very abnormal\textemdash{}see
discussion in Supplementary Information (SI)\textemdash{}both in the
sign of the dependence and the greater amplitude ($\partial Eg/\partial\epsilon$
= 12.3\textendash{}29.7 eV, see Fig. \ref{fig:BLA+gap}), indicating
a different nature of the effect.

Thus, we start with an exploration of the effect of strain on the
Peierls transition. We use a simple analytical model to explain why
tension amplifies the Peierls effect. Then we investigate through
first-principles Born--Oppenheimer potential energy surface calculations
how the atomic vibrations manifest themselves in equilibrium and under
strain. We confirm that the ZPV level lies well above the Peierls
transition energy, and the instability is eliminated. As this energy
is increased by tension, we arrive at the prediction of a strain threshold
whereupon carbyne switches from the ZPV-- to Peierls instability--dominated
regime, accompanied by a sharp change in the electrical conductivity.
Our first-principles numerical analysis validates this prediction,
placing this transition around 3\% strain. Besides carbyne, our findings
naturally suggest that this novel physical effect must also be observable
in other one-dimensional systems exhibiting Peierls behavior, such
as conducting polymers, charge-density-wave materials, or carbon nanotubes.\cite{1992mintmirearefullerene}

Our first-principles data on the amplification of the Peierls instability
under tension are presented in \textbf{Fig.~\ref{fig:BLA+gap}}.
The Peierls distortion is extremely sensitive to the electronic exchange
interaction and, in particular, is poorly described by regular density
functional theory,\cite{2006jacqueminassessment} therefore we used
the HSE06 hybrid functional that includes exact exchange.\cite{2003heydhybridfunctionals,2006heyderratum}
All calculations were also repeated with the PBE generalized-gradient
functional,\cite{1996perdewgeneralized,1997perdewgeneralized} showing
complete qualitative agreement (see SI). Circles and squares represent
the BLA (left axis) and the band gap $E_{g}$ (right axis). The Peierls
transition energy $E_{P}$---energy difference between polyyne and
cumulene structures---is shown in the inset. The striking increase
of all quantities observed so pronouncedly in \textbf{Fig.~\ref{fig:BLA+gap}}
calls for an explanation. 

One can understand the behavior of Peierls transition under strain
using a simple and general model as follows. The lowering of electronic
energy in the Peierls effect is $E_{e}\propto a\left|V_{k}\right|^{2}\log\left|V_{k}\right|/k$,
where $k$ is the Brillouin zone edge ($k=\nicefrac{\pi}{a}$) and
$V_{k}$ is the corresponding Fourier component of the lattice potential.
\cite{2009mihalysolidstate} To determine $E_{e}\left(a\right)$,
some definite form of the $V_{k}\left(a\right)$ dependence is needed.
We use a simple Kronig--Penney type model with two delta-function
atoms per period $2a$ with coordinates $x_{1}=0$ and $x_{2}=\left(a-b\right)/2$,
where $b$ is the BLA (we assume $b>0$). The potential in the unit
cell is $V\left(x\right)=-\delta\left(x-x_{1}\right)-\delta\left(x-x_{2}\right)$,
and then $\left|V_{k}\right|=2\left|\sinh\left(\pi b/2a\right)\right|=\pi\left|b\right|/a+O\left(\left|b\right|^{3}/a^{3}\right)$.
The electronic contribution is counteracted by the lattice deformation
energy, $E_{l}=\nicefrac{1}{2}Cb^{2}+O\left(b^{4}\right)$. The value
of $b$ must minimize the total energy, $\partial\left(E_{e}+E_{l}\right)/\partial b=-\pi b-\pi b\log\left(\pi\frac{b}{a}\right)+Cb=0$,
and we have the answer as $b=\frac{a}{\pi}\exp\left(-\frac{1}{2}-\frac{C}{2\pi}\right)$.
The derivative $db/da$ is positive when $2\pi-a\frac{dC}{da}>0$,
which shows that the Peierls dimerization is increased by tension
unless the restoring force constant increases with strain (at least
as fast as $C=C_{0}+2\pi\log a$---and usually the trend is opposite).

To demonstrate the excellent agreement of our very simple model with
the DFT calculations, we have fitted it to our DFT data (\textbf{Fig.~\ref{fig:BLA+gap}}).
The fitting was done assuming simple linearly decreasing stiffness,
$C\approx C_{0}+D\left(a_{0}-a\right)$, and used only the data points
in the {[}--5\%,~5\%{]} strain interval. The fitting was performed
independently for the BLA and $E_{g}$, and the former fit was used
to compute the line for $E_{P}$.

Up until now, we have confined ourselves to the static picture of
classical carbon atoms. Within this picture, only the polyyne form
of carbyne is stable, and stretching it produces merely \emph{quantitative}
changes in the band gap---a phenomenon known in solid state physics,
even though its origin here is unusual. However, as we demonstrate
below, the quantized nuclear motion changes the picture dramatically.
If ZPV can overpower the Peierls instability in freely-suspended unloaded
chain and restore the symmetric metallic structure, then one is led
to expect that at some critical strain, the stronger Peierls instability
may reverse this phenomenon---causing a \emph{qualitative} change
in properties. Our calculations indeed confirm this.

For the vibrational analysis we computed the Born--Oppenheimer potential
energy surface (PES) profiles along the BLA coordinate, $E_{a}\left(b\right)$,
as the lattice constant $a$ was increased from its equilibrium value.
Because of the extreme anharmonicity of the  $\omega$-shaped PES
profiles, we had to turn to numerical calculations of the vibrational
levels at each strain value. We utilized the Fourier grid Hamiltonian
method \cite{1989marstonthefourier,johnsoniiifourier} which had been
previously used to study the ZPV effect on Peierls transition in unstrained
polyacetylene.\cite{2013hudsonbondalternation} Following Refs.\cite{2005tongayatomicand,2013hudsonbondalternation}
we assume that the longitudinal vibrations can be decoupled from the
two transverse phonons and treated in isolation, which is justified
by the non-negligible bending stiffness of carbyne.\cite{2009ravagnaneffectof,2011hubending,2012castellimechanical}
(Indeed, the inclusion of transverse vibrations and dispersion does
not change our conclusions, as detailed in the SI.) We used a uniform
grid of 250 points in $-0.4\text{ \AA}\le b\le0.4\text{ \AA}$, using
interpolation of DFT-computed values of energy, and 6.005 a.m.u.---the
reduced mass of two \textsuperscript{}C atoms---for the oscillator
mass.
\begin{figure}
\putat{-12}{15}{\begin{sideways}\textsf{energy (meV)}\end{sideways}}\includegraphics[width=0.46\columnwidth]{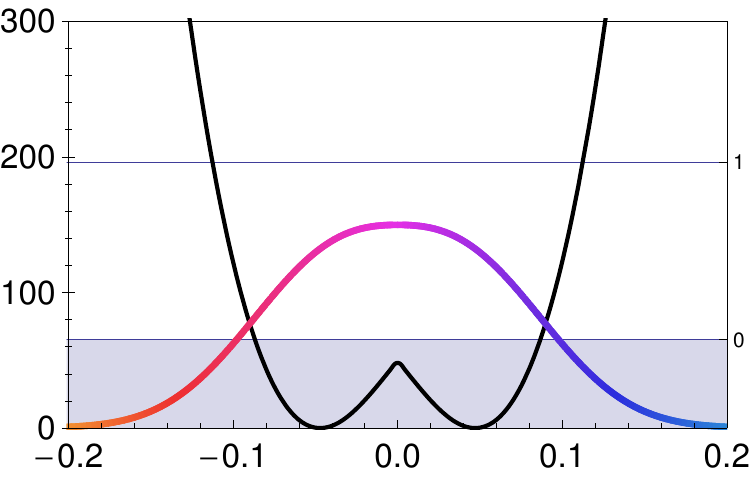}\putat{-100}{60}{$\textsf{0\%}$}\hfill~\includegraphics[width=0.52\columnwidth]{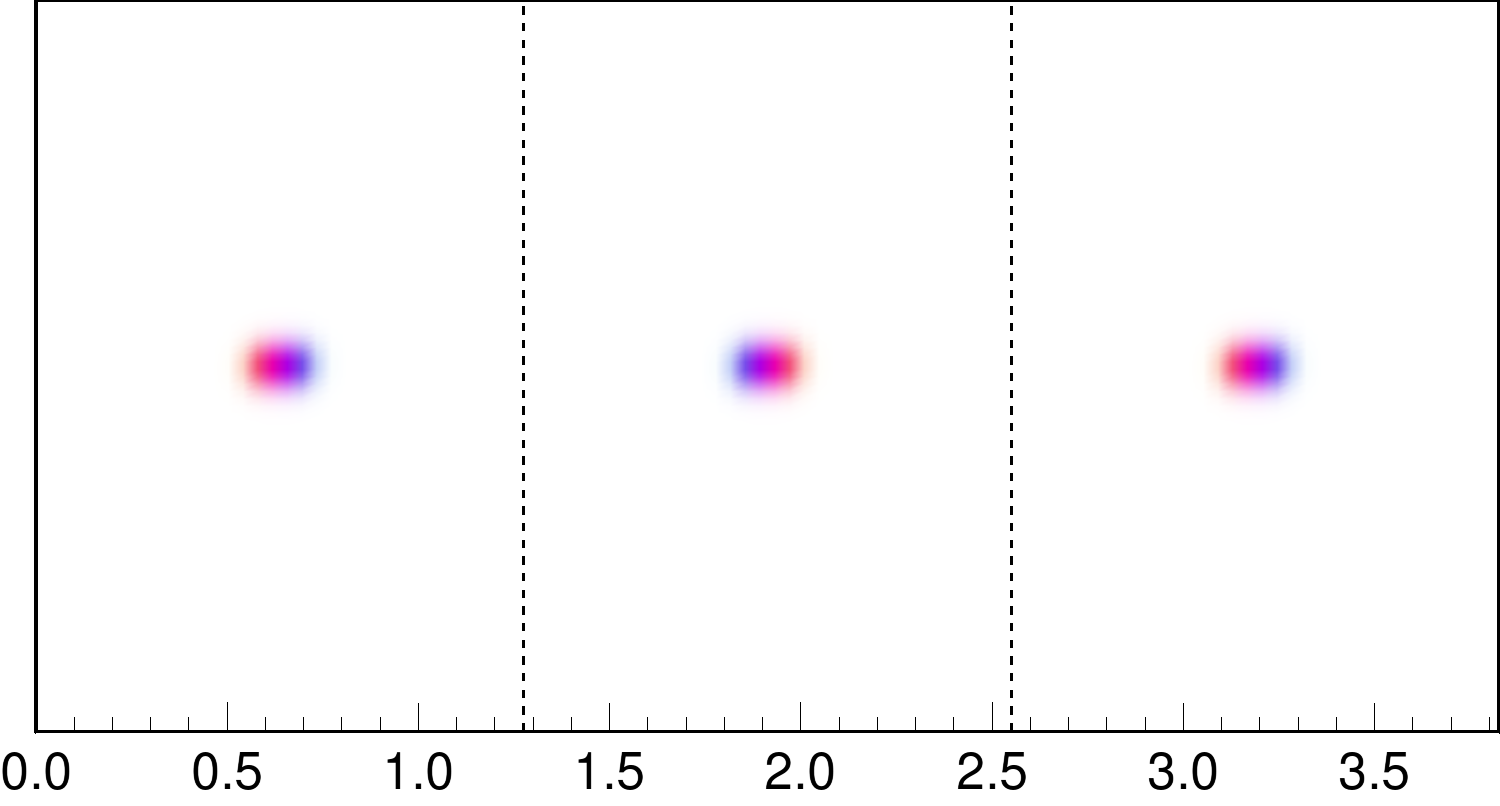}

\putat{-12}{15}{\begin{sideways}\textsf{energy (meV)}\end{sideways}}\includegraphics[width=0.46\columnwidth]{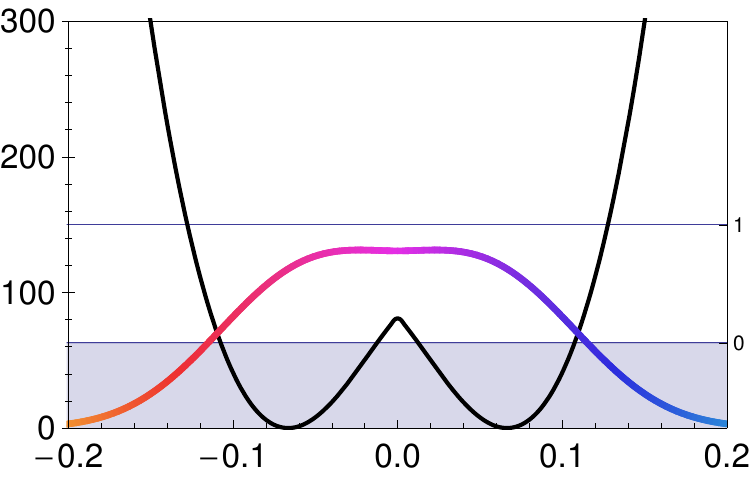}\putat{-102}{60}{$\textsf{3\%}$}\hfill~\includegraphics[width=0.52\columnwidth]{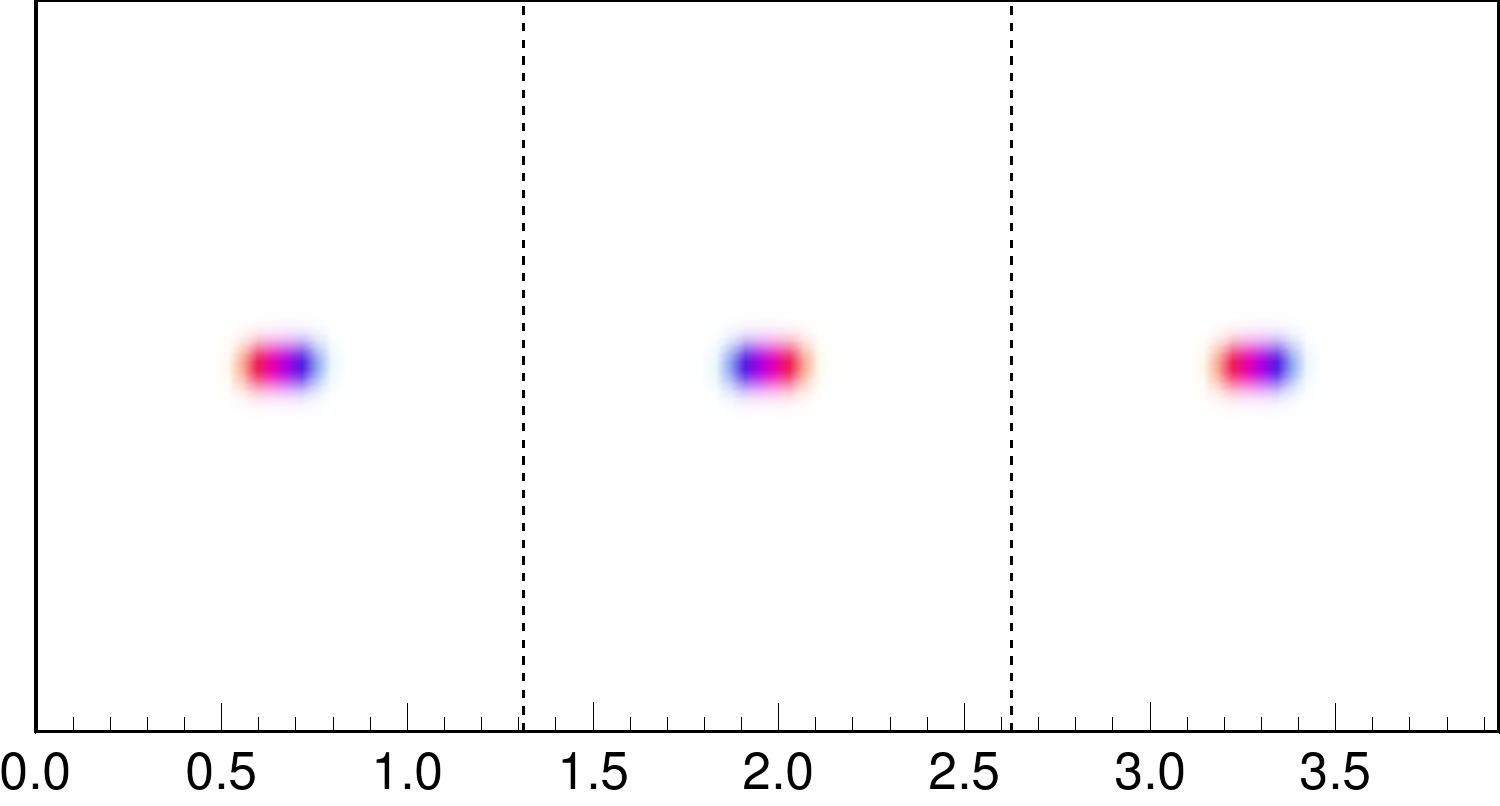}

\putat{-12}{15}{\begin{sideways}\textsf{energy (meV)}\end{sideways}}\includegraphics[width=0.46\columnwidth]{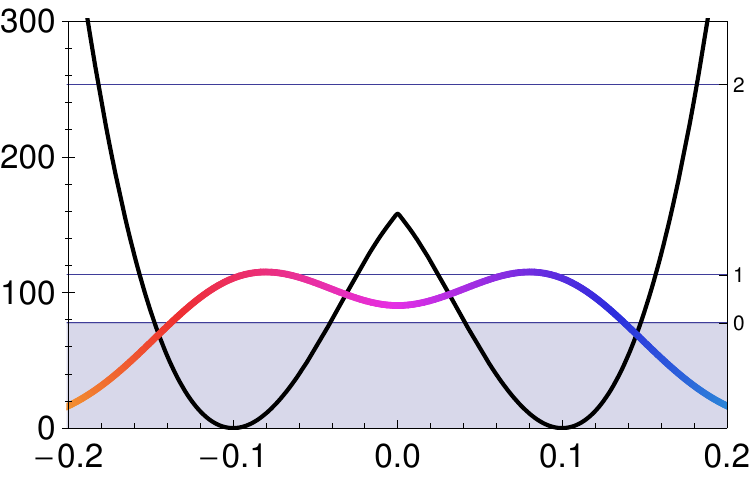}\putat{-97}{60}{$\textsf{7\%}$}\hfill~\includegraphics[width=0.52\columnwidth]{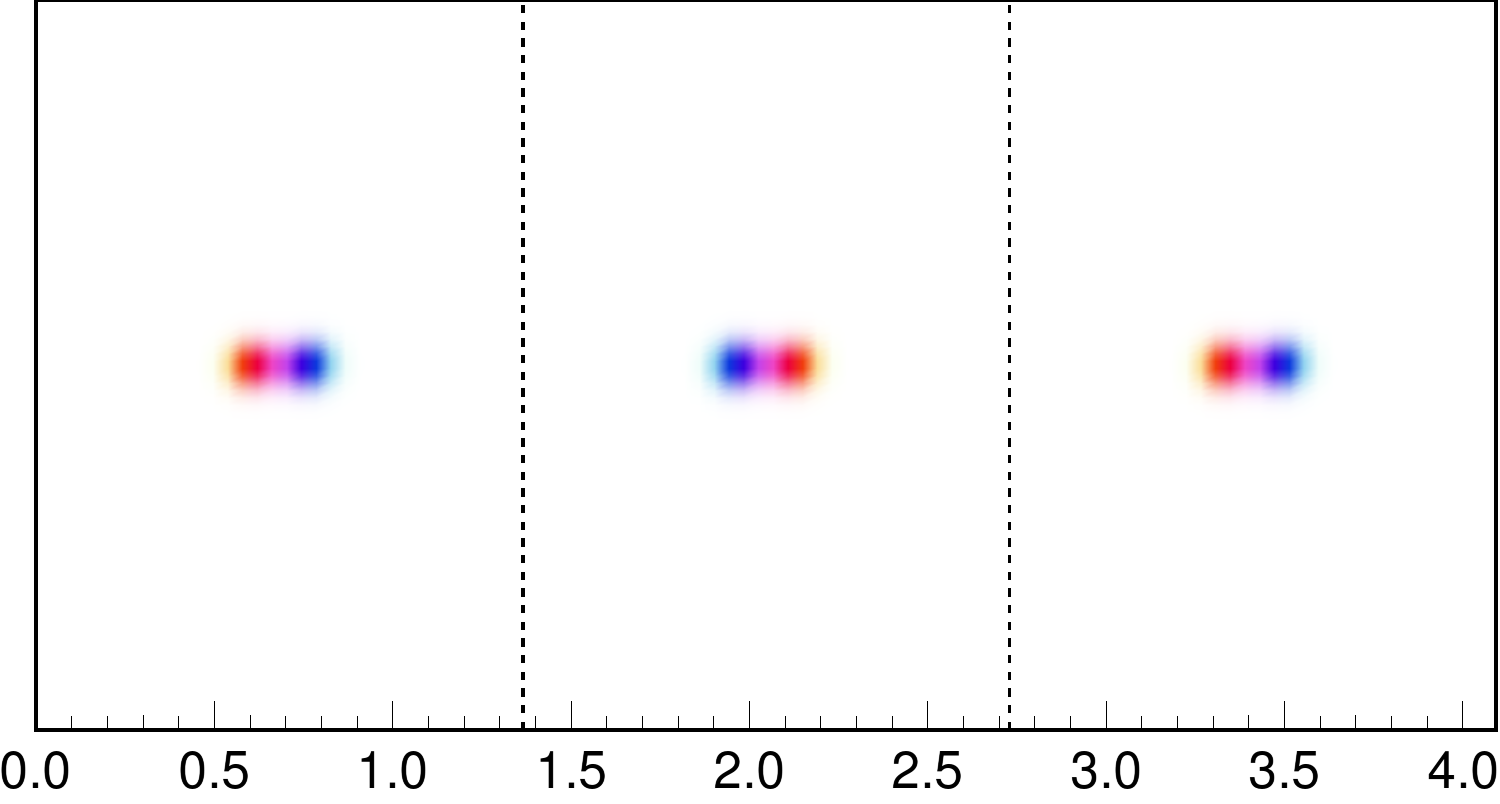}

\putat{-12}{15}{\begin{sideways}\textsf{energy (meV)}\end{sideways}}\includegraphics[width=0.46\columnwidth]{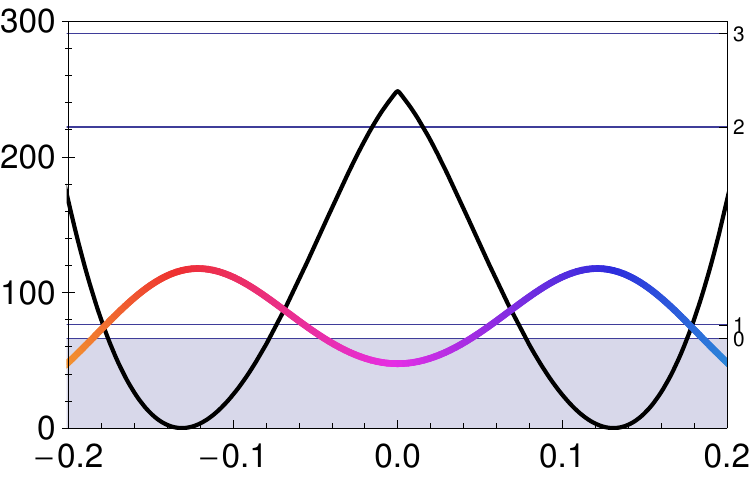}\putat{-100}{59}{$\textsf{10\%}$}\putat{-63}{-8}{$\textsf{\scriptsize{BLA (\AA)}}$}\hfill~\includegraphics[width=0.52\columnwidth]{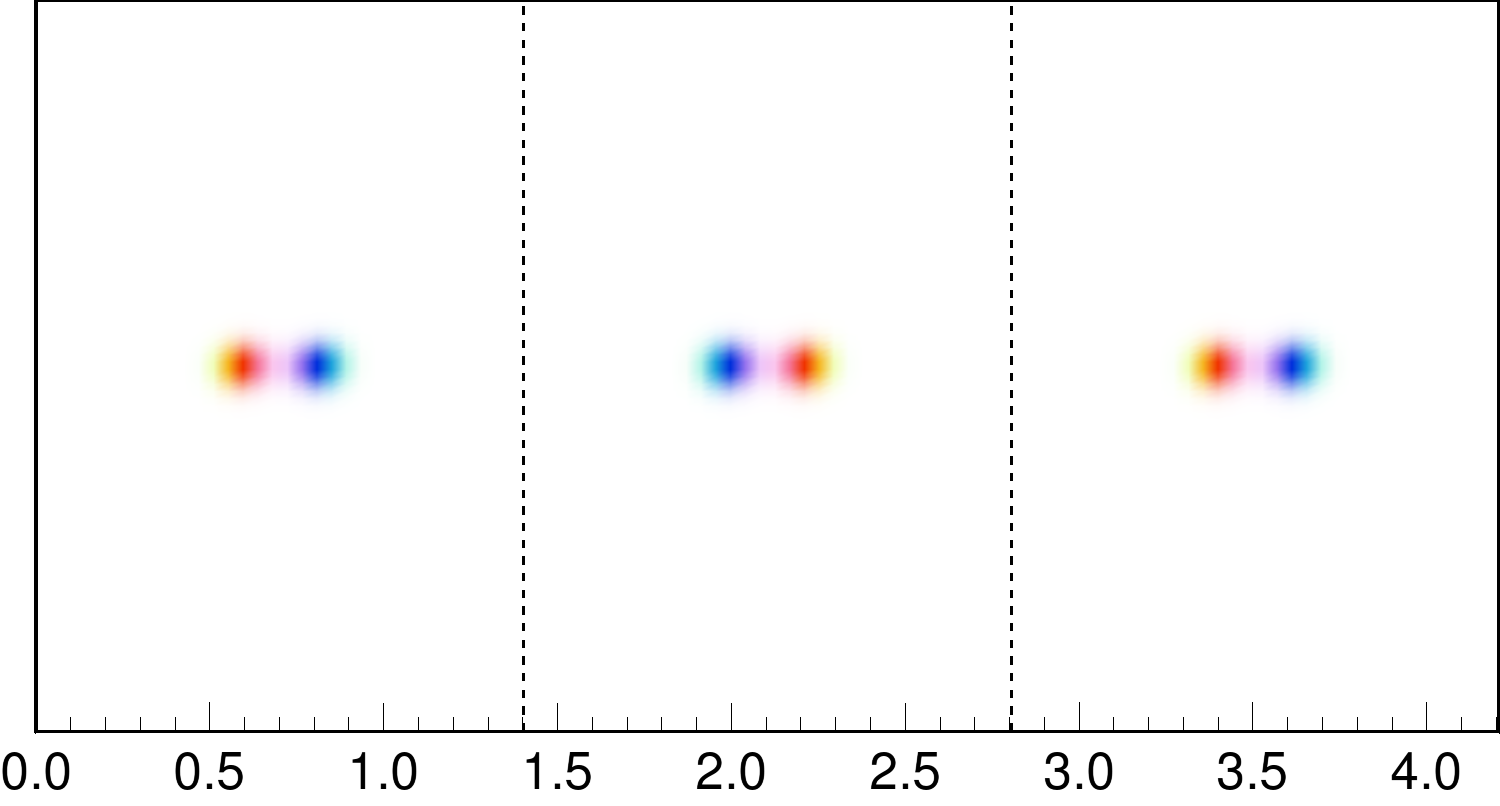}\putat{-82}{-8}{$\textsf{\scriptsize{position (\AA)}}$}

\bigskip

\caption{\textbf{Zero-point vibrations under strain. (left)} Evolution of the
vibrational structure of carbyne with strain. The $\omega$-shaped
lines show the \textbf{potential energy} as a function of BLA. Horizontal
lines show the vibrational levels, the shaded region is below the
\textbf{\textcolor{green}{ZPV level}}. \textbf{(right)} Real-space
atomic density distributions based on ZPV wavefunctions (three one-atom
unit cells shown, color indicates which part of the BLA wavefunction
the density comes from). \label{fig:levels}}
\end{figure}

The results are presented in \textbf{Fig.~\ref{fig:levels}}, which
is the central figure of this paper. The black curves show the $\omega$-shaped
PES as computed using the HSE06 functional, blue horizontal lines
denote the computed vibrational levels with filling below the zero
level, and the red-to-blue curves plot the ZPV wavefunction. The right
panels show reconstructed real-space densities of nuclei at the respective
strain values. The profile along the horizontal axis is the squared
vibrational wavefunctions with an arbitrary transverse broadening.
Colors represent which values of BLA (positive or negative in the
left-hand side plots) correspond to the position.

Clearly, the mechanically relaxed system has a single wavefunction
maximum with a width exceeding the BLA,\cite{1992mckenzieeffectof}
corresponding to the cumulene structure of carbyne with lattice parameter
$\nicefrac{a}{2}$. Thus, ZPV stabilize the cumulene form even though
it is a maximum on the PES. In accordance with the above discussion,
the central PES maximum ($E_{P}$) rises with strain, and at 3\%,
the wavefunction develops a slight dip in the center, resulting in
a picture of `elongated' atoms. At 7\% strain, the ground-state wavefunction
is clearly separated in two blobs, and the first excited level approaches
the ground state, indicative of the transition to the double-well
potential--like regime. Under further stretching (10\%), the first
excited level is well below room-temperature $k_{B}T$ of the ground
state, as in two independent potential wells. The second level dives
below the Peierls barrier, and the third begins to approach it (at
15\% strain these two also become degenerate; data not shown). The
joint evolution of Peierls barrier and vibrational levels under stretching
is further illustrated in the SI (Fig. S2).

\begin{figure}
\textsf{temperature (K)}\hfill~

\hfill~\putat{125}{-8}{$\textsf{strain (\%)}$}\includegraphics[width=1\columnwidth]{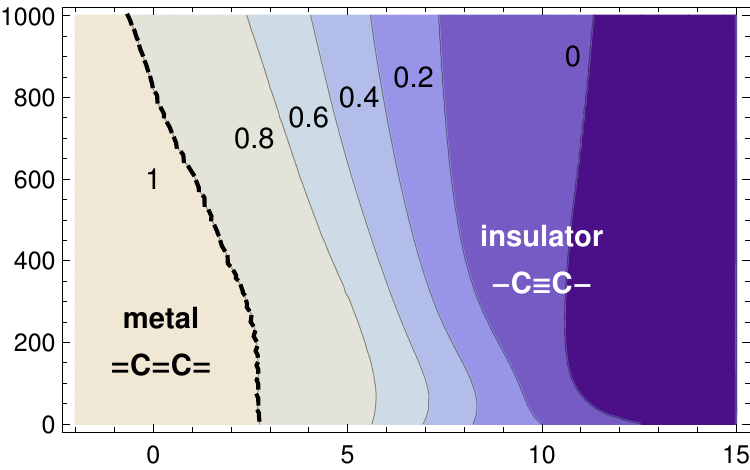}

\smallskip

\caption{\textbf{Strain--temperature ``phase diagram'' of the polyyne--cumulene
transition. }The order parameter $R$ (see the text) measures the
shape of the nuclear density distribution, with one maximum (cumulene)
at $R=1$ and two (polyyne) at $R<1$.\label{fig:vib-phase}}
\end{figure}
The above-described sequence of events has a profound consequence.
As the strain is increased, carbyne transitions from the conducting
electronic behavior of cumulene to the nonconductive polyyne structure
with its sizable band gap (see \textbf{Fig.~\ref{fig:BLA+gap}}).
One can go a step further and compute a strain\textendash{}temperature
``phase diagram\textquotedblright{} for carbyne. First, the temperature-dependent
nuclear densities $n_{T}\left(b\right)$ are computed using Bose statistics
from squared wavefunctions. A rising temperature and strain-induced
depression of vibrational levels start to populate the first excited
states which have small amplitude near $b=0$, and $n_{T}\left(b\right)$
becomes wider until its central maximum splits in two. To determine
the polyynic or cumulenic character, for the role of order parameter
we used a measure of the distribution shape: the ratio $R=n_{T}\left(0\right)/\max_{b}\left(n_{T}\left(b\right)\right)$,
which is 1 always when there is a single central minimum (pure cumulene)
and less than 1 when the symmetry is broken. \textbf{Fig.~\ref{fig:vib-phase}}
shows a contour map of $R$. The region to the left of the $R=1$
contour corresponds to pure cumulene. At low temperatures, the threshold
strain is 3\% and it decreases with temperature starting from about
200 K, until at 900 K, the cumulene form becomes unstable (unless
compressed, which is hard to picture with carbyne but may be feasible
for other systems such as carbon nanotubes). Note that in the canonical
Peierls effect, high temperatures tend to restore symmetry. That is
an electronic excitation effect, and it should be negligible given
polyyne\textquoteright{}s wide band gap. (Further, this effect could
only result in a shift of the transition to larger strains.) Thus,
once again we see that the inclusion of nuclear degrees of freedom
in the picture has led to unexpected qualitative changes of behavior. 

Besides the strain and temperature, the composition of carbyne provides
offers another degree of freedom. With \textsuperscript{13}C instead
of \textsuperscript{12}C, all levels would experience a slight red-shift,
moving the cumulene stability boundary closer to the origin. Formally,
lighter C isotopes (had such existed) or lighter elements (assuming
constant bond stiffness) would have moved the boundary up and to the
right.

\begin{figure}
\bigskip

\putat{-10}{28}{\begin{sideways}$\textsf{bond length (\AA)}$\end{sideways}}\putat{85}{-8}{$\textsf{bond no. in R--C\textsubscript{12}--R}$}\putat{82}{130}{$\textsf{bond no. in R--C\textsubscript{100}--R}$}\includegraphics[width=0.95\columnwidth]{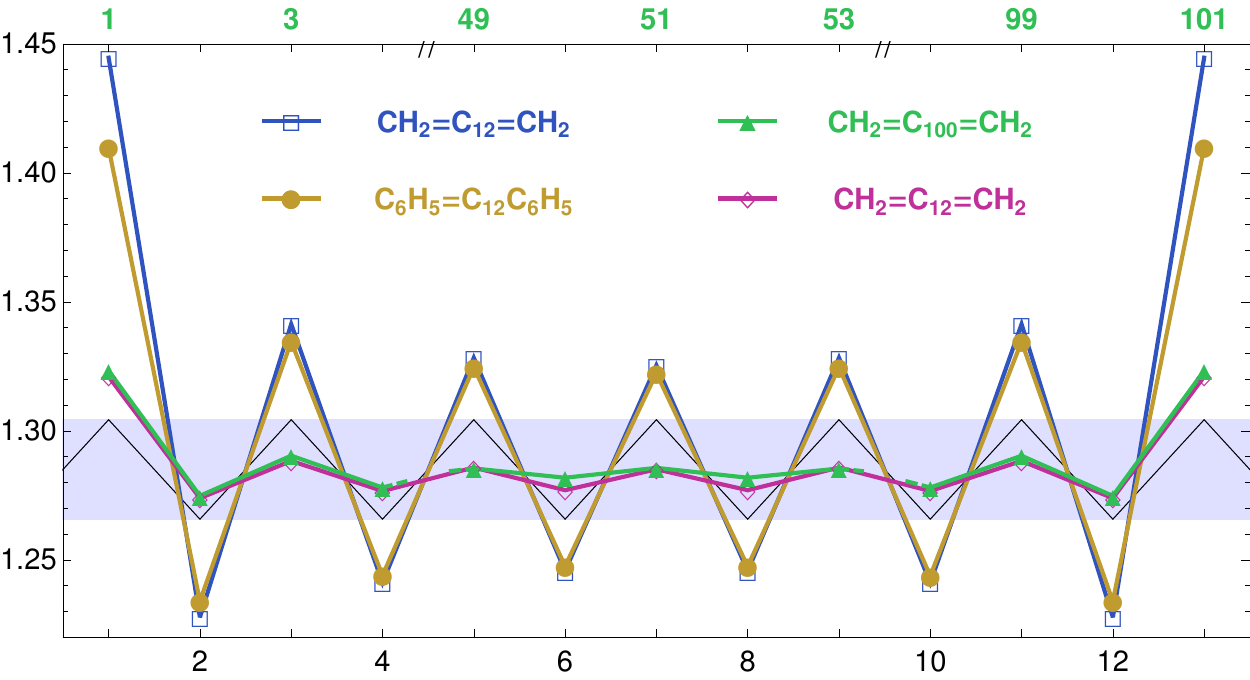}\hfill\bigskip

\includegraphics[width=1\columnwidth]{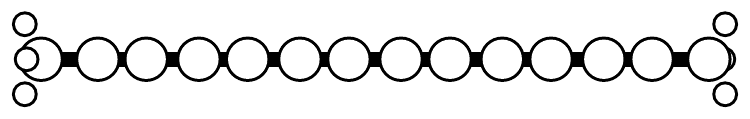}

\caption{\textbf{Effect of endgroups on carbyne BLA.} Termination of the chain
with \emph{sp}\textsuperscript{3} groups (methyl\textbf{ }\textbf{\textcolor{blue}{--CH\textsubscript{3}}}
and phenyl\textbf{ }\textbf{\textcolor{yellow}{--C\textsubscript{6}H\textsubscript{5}}})
increases the BLA, whereas the \emph{sp}\textsuperscript{2} methylene
group (\textbf{\textcolor{magenta}{=CH\textsubscript{2}}}) decreases
it, as compared to an unterminated infinite chain (black dashed line).
The data for \textbf{\textcolor{magenta}{12-}} and \textbf{\textcolor{green}{100-atom}}
chains show a weak decrease of BLA with chain length away from the
ends with methylene termination.\label{fig:Endgroups}}
\end{figure}

So far our calculations treated carbyne as an infinite chain. The
PES of infinite carbyne contains pairs of minima and is invariant
under translation by one-half the unit cell of polyyne, which corresponds
to the exchange of triple and single bonds. However, the ends of the
carbyne chain, inevitable in any real experiment, will impose some
boundary conditions that can break this degeneracy. (Note that our
predictions should still apply directly to carbyne rings.) As a simple
example, the two \emph{finite} structures described by structural
formulae 
\[
\text{H}-\text{C}\left(\equiv\text{C}-\text{C}\equiv\right)_{n}\text{C}-\text{H}
\]
and
\[
\text{H}\equiv\text{C}\left(-\text{C}\equiv\text{C}-\right)_{n}\text{C}\equiv\text{H}
\]
are very nondegenerate, because H cannot even form stable triple bonds
with C. On the other hand, the use of groups that interface with the
chain through a double bond, such as methylene =CH\textsubscript{2},
could favor the cumulene structure and stabilize the symmetry. The
effect of end-groups will clearly have a different strength for different
moieties, and a comprehensive investigation is an extensive task (further
aggravated by the complexity of collective atomic motions in such
many-atom systems) that is beyond the scope of the present study.
However, simply looking at how the BLA depends on the end-group type
in finite carbyne fragments already provides a useful glance into
the issue. 

The results of our finite-chain calculations with different terminating
groups are presented in \textbf{Fig.~\ref{fig:Endgroups}}. The strip
from 1.27 to 1.30 \AA ~denotes the bond lengths in the ideal infinite
polyyne chain (ignoring the ZPV). The bond lengths in the \emph{sp}\textsuperscript{3}-terminated
systems (--CH\textsubscript{3}, --C\textsubscript{6}H\textsubscript{5})
fall outside this strip, showing the enhanced polyyne-like character
of the systems. On the other hand, \emph{sp}\textsuperscript{2}-type
=CH\textsubscript{2} termination reduces the BLA effectively to zero
(< 0.01 \AA ~for a 12-atom chain and < 0.005 \AA ~in the middle of a 100-atom
chain), suggesting that this type of passivation may even be used
to shift the transition boundary towards higher strains/temperatures.
(Indeed, \emph{sp}\textsuperscript{2}-type groups are preferable
from the electronics applications standpoint, providing metallic 'leads'
at the ends of the chain.) We see that in experimental realizations,
the switching effect may be destroyed by improper chemical termination
of carbyne chains, and thus, precise control over both the tension
and chemistry is essential.

In conclusion, we have explored how the Peierls instability of linear
carbon chains is enhanced by stretching using first-principles calculations,
and explained it with a simple analytical model. At zero strain, the
Peierls instability in carbyne is too weak compared to quantum zero-point
vibrations, and the stable structure corresponds to the metallic cumulene
rather than insulating polyyne---despite the lower potential energy
of the latter structure. However, the enhancement of Peierls instability
results in the reversal of this effect, and carbyne starts to demonstrate
the polyyne-like character of the nuclear wavefunction, switching
from the conducting to insulating behavior around 3\% strain. This
effect naturally applies to other one-dimensional Peierls systems.
Our findings are both important as an interesting new fundamental
physical effect and highly practically relevant for the science of
conducting polymers, charge- and spin-density-wave materials, and
electromechanical applications. Although carbyne remains an exotic,
its unique properties will continue to fuel the effort to achieve
its synthesis in practically useful quantities.

The computations were performed using the \textsc{VASP} code \cite{1996kresseefficient}
with projector-augmented wave basis sets,\cite{1994blochlprojector,1999kressefromultrasoft}
using the Data Analysis and Visualization Cyberinfrastructure funded
by NSF under Grant OCI-0959097.

\bibliographystyle{naturemag}
\bibliography{Peierls}

\end{document}